\documentclass{PoS}

\usepackage{graphicx}
\usepackage{amsmath}
\usepackage{amssymb}
\usepackage[sort&compress,numbers]{natbib}
\usepackage{horowitzarXiv}

\newcommand{\TITLE}{Qualitative and Quantitative Jet Physics in Pb + Pb at LHC}

\title{\TITLE}

\ShortTitle{LHC Jet Physics}

\author{\speaker{W.\ A.\ Horowitz}\\
        Department of Physics, University of Cape Town, Private Bag X3, Rondebosch 7701, South Africa\\
        Department of Physics, The Ohio State University, 191 W.\ Woodruff Ave., Columbus, OH 43210, USA\\
        E-mail: \email{wa.horowitz@uct.ac.za}}

\abstract{The momentum dependence of the exciting new Pb + Pb data from LHC qualitatively suggests a perturbative picture interpretation for the energy loss of \highpt particles, but conclusions are difficult to draw due to the lack of 1) a quantitative theoretical calculation constrained by current high precision RHIC data and 2) a lack of control p + p and p + Pb data at LHC.  Future measurements of identified heavy quark suppression will provide a novel qualitative tool for determining the dominant physics of the quark-gluon plasma created at RHIC and LHC.}

\FullConference{Workshop on Discovery Physics at the LHC -Kruger 2010\\
		December 05-10, 2010\\
		Kruger National Park, Mpumalanga, South Africa}

\begin{document}

\section{Introduction}
\begin{figure}[!htbp]
\centering
$\begin{array}{cc}
\includegraphics[width=.48 \columnwidth]{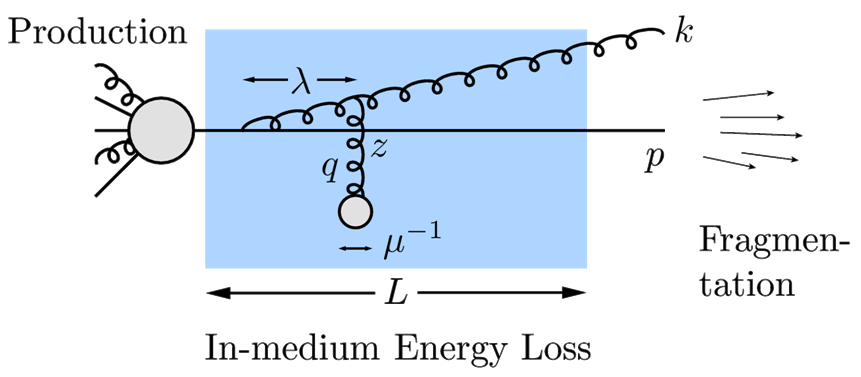} &
\includegraphics[width=.48 \columnwidth]{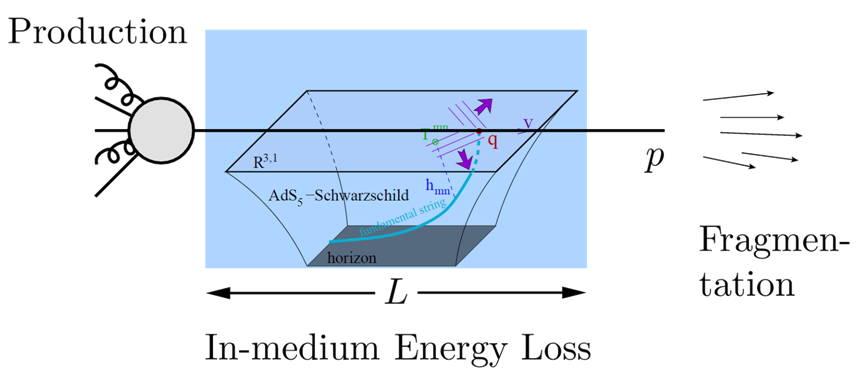} \\[-10pt]
\hspace{-2in}\mathrm{(a)} & \hspace{-2in}\mathrm{(b)}
\end{array}$
\caption{\label{fig:diagram}
Cartoon of the production, in-medium energy loss, and fragmentation processes that may occur using (a) pQCD for in-medium energy loss or (b) AdS/CFT for in-medium energy loss for a \highpt parton produced in a heavy ion collision.  Figures adapted from \cite{Horowitz:2009eb} and \cite{Friess:2006fk}.
}
\end{figure}

The QCD Lagrangian is known and well verified by particle experiments over many orders of magnitude  \cite{Cacciari:1998it,Jager:2002xm,deFlorian:2002az}.  But just as the collective behavior of electrically charged objects---take for instance the phase diagram of water---is far from obvious given the QED Lagrangian, the bulk dynamics of QCD are not well understood \cite{LRP}.  With the Relativistic Heavy Ion Collider (RHIC), the Large Hadron Collider (LHC), and the soon-to-be-running Facility for Antiproton and Ion Research (FAIR) facilities the world scientific community has a novel opportunity to experimentally explore the state of the universe a few microseconds after the Big Bang and the phase transition from the usual, confining hadronic matter to deconfined quarks and gluons by colliding heavy nuclei, such as gold and lead, at near the speed of light.  In particular rare high momentum partons, those that produce jets of particles, provide the most direct probe of the fundamental degrees of freedom in the new phase of QCD matter created at a few times the transition temperature, $T_c$, at these cutting-edge facilities \cite{Gyulassy:2004zy,Wiedemann:2009sh,Majumder:2010qh}.  \fig{fig:diagram} shows a cartoon of the energy loss of these \highpt probes as they propagate through a deconfined plasma of quarks and gluons in (a) completely perturbative---calculable via the methods of perturbative QCD (pQCD)---and (b) completely nonperturbative---calculable via the methods of AdS/CFT---pictures.  The goal, then, is to compare theoretical predictions to data to determine the properties of the quark-gluon plasma created in heavy ion collisions.

\section{Results}

The suppression of \highpt particles in heavy ion collisions is usefully expressed in terms of \raacomma: the ratio of measured yield in nucleus-nucleus (A + A) collisions divided by the yield in p + p collisions scaled by the expected number of p + p collisions in an A + A collision.  Extracting information from \raacomma, and suppression observables in general, is made nontrivial by the many physics processes---some of which are interesting in their own right---that are effectively integrated out.  Of greatest import for our discussion is the initial distribution of \highpt quarks and gluons in both coordinate and momentum space.  At RHIC, reference spectra from p + p collisions at the same $\sqrt{s}$ as in Au + Au collisions \cite{Adams:2005ph,Adare:2011vy} allow for a precise measurement of \raa \cite{Adams:2005ph,Adare:2008cg} and constrain the initial quark and gluon spectra; there is also abundant evidence from the d + Au collision data that at midrapidity the initial \highpt partonic spectrum is not significantly modified from that in p + p collisions \cite{Adams:2003im,Adler:2006wg}, and direct photon measurements reassure us that the binary distribution from a Glauber model approximates well the number of hard proton-proton-like collisions in a heavy ion event \cite{Adler:2005ig}.  

\begin{figure}[!htbp]
\centering
$\begin{array}{ccc}
\includegraphics[width=.3 \columnwidth]{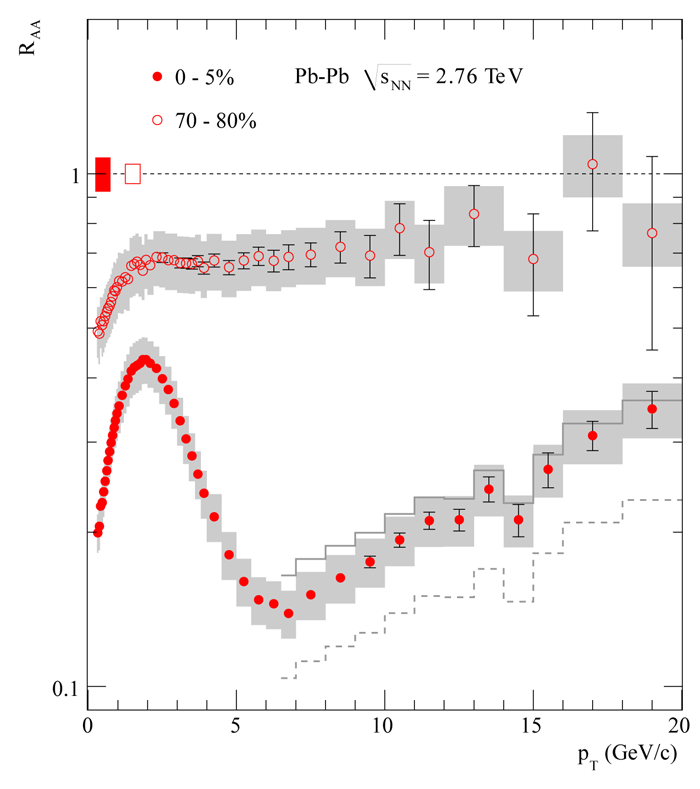} & \hspace{1in} &
\includegraphics[width=.35 \columnwidth]{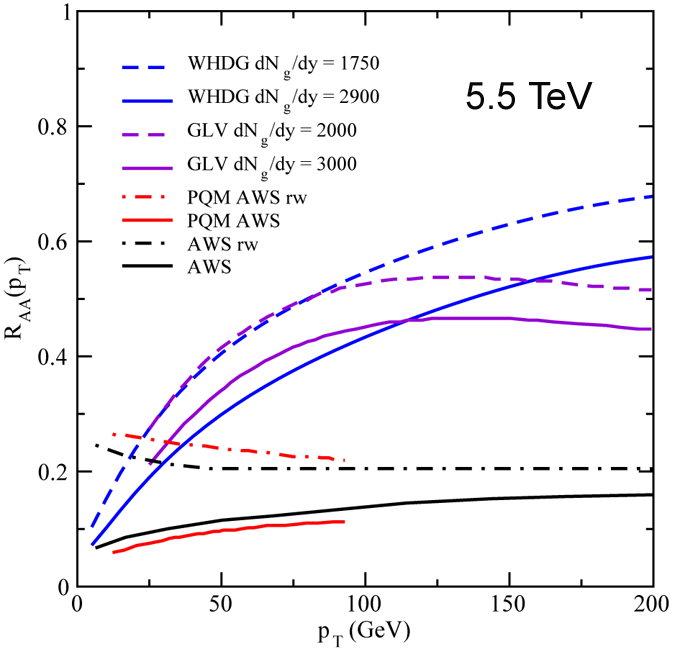} \\[-10pt]
\hspace{-2in}\mathrm{(a)} & \hspace{1in} & \hspace{-2in}\mathrm{(b)}
\end{array}$
\caption{\label{fig:RAA}
(a) Measured $R_{AA}$ of charged hadrons at 2.76 TeV at LHC \cite{Aamodt:2010jd}.  (b) Predictions from the WHDG energy loss model \cite{Wicks:2005gt} for the $R_{AA}$ of pions at 5.5 TeV \cite{Horowitz:2007nq}.
}
\end{figure}

\begin{figure}[!htbp]
\centering
$\begin{array}{ccc}
\includegraphics[width=.4 \columnwidth]{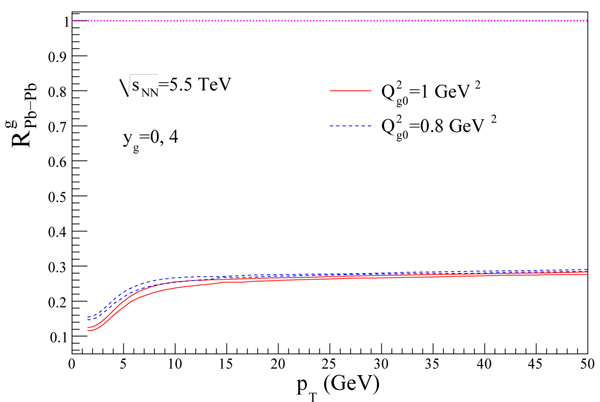} & 
&
\includegraphics[width=.55 \columnwidth]{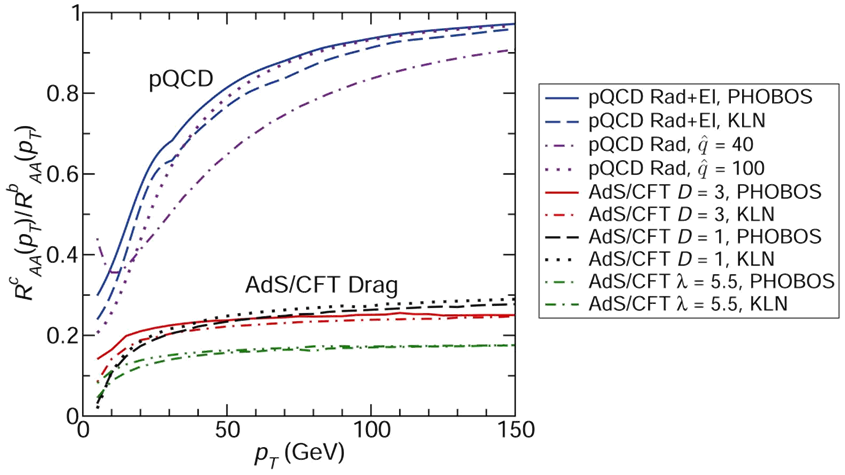} \\[-10pt]
\hspace{-2in}\mathrm{(a)} & 
& \hspace{-2in}\mathrm{(b)}
\end{array}$
\caption{\label{fig:CGC}
(a) Predictions of the suppression of gluons in top energy Pb + Pb LHC collisions from initial state effects \emph{only} \cite{Albacete:2010bs}. (b) The ratio of charm to bottom $R_{AA}$ should provide a clear indication of the dominant physics of \highpt energy loss \cite{Horowitz:2007su}.}
\end{figure}

These measurements do not yet exist at LHC, critically limiting the precision of---and our ability to interpret---the recent \raa results \cite{Aamodt:2010jd}, shown in \fig{fig:RAA} (a).  First, there is no reference p + p data, which leads to about a factor of 2 $p_T$-dependent systematic uncertainty in the normalization of the reported \raacomma.  Second, there is no ``control measurement'' from a p + Pb run.  This control measurement is crucial as there are predictions of a significant \emph{initial state} suppression of \highpt particles due to unitarizing effects in the initial gluon parton distribution function \cite{Albacete:2010bs}; see \fig{fig:CGC} (a).  Although not shown in that work, these effects should be $x$ and $Q^2$ dependent (the effects should disappear by $\eqnpt\sim25-50$ GeV/c) naturally introducing a rise in \raa as a function of \ptcomma, perhaps similar to that seen currently in the data.  The ATLAS measurement \cite{Collaboration:2010px} consistent with a lack of suppression in its 38 Z boson candidates is encouragingly suggestive that the binary scaling seen at RHIC also holds at LHC, although the result is currently inconclusive due to the large systematic uncertainty and lack of statistics.  

There is currently debate over even a qualitative understanding of the energy loss mechanism(s) in a QGP \cite{Horowitz:2007su,Marquet:2009eq,Horowitz:2010yi,Jia:2011pi}.  Current pictures assume that either all interactions are strongly-coupled, some are strong and some are weak, or that all interactions are weak (see \cite{Horowitz:2007su,Majumder:2010qh} and references therein).  It is worth noting, though, that none of the purely perturbative approaches quantitatively describes \emph{simultaneously} any two single particle \highpt observables at RHIC.  

Nevertheless, the new ALICE data \cite{Aamodt:2010jd} shows a feature that appears strikingly similar to what one would expect from pQCD-based energy loss: \raa rises significantly as a function of \ptcomma.  One expects a rise as a function of momentum as the fractional energy loss of a \highpt parton goes as $\epsilon\sim\log(\eqnpt)/\eqnpt$ \cite{Gyulassy:2000fs}, where the final momentum, $\eqnpt^f$, is related to the initial momentum, $\eqnpt^i$, by $\eqnpt^f=(1-\epsilon)\eqnpt^i$.  If particle production is well approximated by a power law, $dN/d\eqnpt\sim\eqnpt^{-n}$, then $\eqnraa \sim \langle (1-\epsilon)^{n-1} \rangle$ \cite{Horowitz:2010dm}.  The suppression at RHIC is flat within the uncertainty of the experiment \cite{Adare:2008cg}.  If the picture of perturbative energy loss is correct, this flatness is due to a coincidental cancellation between 1) the fraction of \highpt gluons to quarks, 2) the hardening of the production spectrum as a function of \ptcomma, and 3) the decrease in energy loss as a function of \ptcomma.  At LHC, the production spectrum is much flatter than at RHIC while energy loss till decreases with \ptcomma; this then would lead to an \raa that increases with \ptcomma.  One can see from \fig{fig:RAA} (b) that that is exactly what we find for the WHDG model of pQCD-based energy loss we study here.  

Furthermore, ALICE reports \cite{Aamodt:2010pb} a factor $\sim2.2$ increase in the central charged particle multiplicity at LHC compared to RHIC.  Assuming that the quark-gluon plasma medium density scales with charged particle multiplicity, it becomes a quantitative question whether an energy loss calculation can simultaneously describe the \raa normalization at both RHIC and LHC.  

An exciting future measurement is of the ratio of suppression patterns of charm and bottom quarks, $R_{AA}^c/R_{AA}^b$, as a function of \ptcomma, shown in \fig{fig:CGC} (b).  One readily notices how the interplay of mass and momentum scales, very different for perturbative and AdS/CFT energy loss calculations, manifests itself: in perturbative calculations the momentum loss per unit time scales as $dp_T/dt \sim -LT^3\log(p_T/M_Q)$ whereas in the AdS/CFT calculations $dp_T/dt \sim -(T^3/M_Q^2)p_T$, where $L$ is the pathlength traversed by the \highpt parton, $T$ is the temperature of the plasma, and $M_Q$ is the mass of the heavy quark.

\section{Conclusions and Outlook}

A new era in heavy ion physics started on November 8$^\mathrm{th}$, 2011 when LHC began colliding lead nuclei at $\sqrt{s}_{NN}$ = 2.76 TeV, an unprecedented center of mass energy.  While the initial \highpt results have qualitative features that appear to reflect the dominance of perturbative physics, one must quantitatively compare theoretical predictions to data at both RHIC and LHC energies simultaneously.  However, even before making this comparison, it is clear that even a qualitative scientific conclusion based on the measured suppression of charged hadrons will be difficult to reach due to the influence of possibly very large, currently experimentally unconstrained initial state effects and to the large systematic uncertainty in the initial spectrum of \highpt particles.  Future measurements from p + p and p + A runs at the same $\sqrt{s}_{NN}$ will greatly improve the ability for the community to reach a scientific consensus; future measurements of heavy quark suppression at LHC will provide a novel qualitative tool for investigating the gross properties of the quark-gluon plasma, such as whether the medium is best described as a strongly coupled fluid or a weakly coupled plasma.

\providecommand{\href}[2]{#2}\begingroup\raggedright\endgroup

\end{document}